\documentstyle[11pt,epsfig]{article}

\begin{document}

\def \cl {{\cal L}}
\def \be {\begin{equation}}
\def \ee {\end{equation}}
\def \bea {\begin{eqnarray}}
\def \eea {\end{eqnarray}}

\def \bi {\bibitem}
\def \ci {\cite}
\def \e {{\rm e}}
\def \o {\omega}
\def \a {\alpha}
\def \n {\nu}
\def \cf {{\cal F}}
\def \L {\Lambda}
\def \x {\xi}
\def \g {\gamma}
\def \del {\partial}
\def \pt {{\rm PT}}
\def \eps {\varepsilon}

\def \p {\pi}
\def \m {\mu}
\def \cs {{\cal S}}
\def \as {{\alpha_s}}

%%%%%%%%%%%%%%%%%%%%%%%%%%%%%%%%%%%%%%%%%%%%%%%%%%%%%%%%%%%%%%%%%%%%%%%%%%%%
%%%%%%%%%%%% Options: preprint*, proceedings, published, (no)hyper*, paper,%
%%%%%%%%%%%%          a4paper*, letterpaper, legalpaper, executivepaper, %%%
%%%%%%%%%%%%          11pt, 12pt*, oneside*, twoside, draft %%%%%%%%%%%%%%%%
%%%%%%%%%%%%%%%%%%%%%%%%%%%%%%%%%%%%%%%%%%%%%%%%%%%%%%%%% *=default %%%%%%%%
%%%%%%%%%%%% \conference{...} %%%%%%%%%%%%%%%%%%%%%%%%%%%%%%%%%%%%%%%%%%%%%%
%%%%%%%%%%%% \title{...} %%%%%%%%%%%%%%%%%%%%%%%%%%%%%%%%%%%%%%%%%%%%%%%%%%%
%%%%%%%%%%%% \author{...\thanks{...}\\...} %%%%%%%%%%%% \email{...} %%%%%%%%
%%%%%%%%%%%% \abstract{...} %%%%%%%%%%%%%%%%%%%%%%%%%%%%%%%%%%%%%%%%%%%%%%%%
%%%%%%%%%%%% \keywords{...} %%%%%%%%%%%%%%%%%%%%%%%%%%%%%%%%%%%%%%%%%%%%%%%%
%%%%%%%%%%%% \preprint{...} %% or \received{...} \accepted{...} \JHEP{...} %
%%%%%%%%%%%%%%%%%%%%%%%%%%%%%%%%%%%%%%%%%%%%%%%%%%%%%%%%%%%%%%%%%%%%%%%%%%%%
%%%%%%%%%%%% \dedicated{...} %%%%%%%%%%%%%%%%%%%%%%%%%%%%%%%%%%%%%%%%%%%%%%%
%%%%%%%%%%%%%%%%%%%%%%%%%%%%%%%%%%%%%%%%%%%%%%%%%%%%%%%%%%%%%%%%%%%%%%%%%%%%
%%%%%%%%%%%% \aknowledgments %%%%%%%%%%%%%%%%%%%%%%%%%%%%%%%%%%%%%%%%%%%%%%%
%%%%%%%%%%%%%%%%%%%%%%%%%%%%%%%%%%%%%%%%%%%%%%%%%%%%%%%%%%%%%%%%%%%%%%%%%%%%
%%%%%%%%%%%% -- No pagestyle formatting. %%%%%%%%%%%%%%%%%%%%%%%%%%%%%%%%%%%
%%%%%%%%%%%% -- No size formatting. %%%%%%%%%%%%%%%%%%%%%%%%%%%%%%%%%%%%%%%%
%%%%%%%%%%%% Your definitions: %%%%%%%%%%% These are mine... :) %%%%%%%%%%%%
%   ... 								   %
\newbox\mybox
\newcommand{\ttbs}{\char'134}           % \backslash for \tt (Nucl.Phys. :)%
\newcommand\fverb{\setbox\mybox=\hbox\bgroup\verb}
\newcommand\fverbdo{\egroup\medskip\noindent\fbox{\unhbox\mybox}\ }
\newcommand\fverbit{\egroup\item[\fbox{\unhbox\mybox}]}

%\font\beeg=cmr17 scaled 1600		% Stylish initials
%\font\beeg=yinit scaled 800
%\newcommand\init[1]{\setbox\mybox=\hbox{{\beeg #1}~}%
%		   \noindent\global\hangindent=\wd\mybox\global\hangafter-2%
%		   \sc\smash{\llap {\lower 13.2pt \box\mybox}}}
%   ...                                                                    %
%%%%%%%%%%%%%%%%%%%%%%%%%%%%%%%%%%%%%%%%%%%%%%%%%%%%%%%%%%%%%%%%%%%%%%%%%%%%

\begin{flushright}
YITP-SB-02-80\\
January 23, 2003
\end{flushright}

\bigskip

\centerline{\Large {\bf Approaching the Final State}}

\smallskip

\centerline{{\Large {\bf in Perturbative QCD}}}

\par\vspace*{8mm}\par

\begin{center}
{\large {\bf
George Sterman}}

\par\vspace*{5mm}\par

{\it 
C.N.\ Yang Institute for Theoretical Physics\\
Stony Brook University, State University of New York\\
Stony Brook, New York 11794 -- 3840, U.S.A.}
\end{center}

\begin{abstract}
Infrared safe differential cross sections, such as
event shape distributions, can be measured
over wide kinematic ranges,
from regions where fixed order calculations are adequate to regions where 
nonperturbative dynamics dominate.
Such observables provide an ideal laboratory for the study
of the transition between weak and strong coupling in quantum
field theory.   This talk begins with some of the fundamentals
of the perturbative description of QCD and
the basis of resummation techniques, followed by a brief
discussion of selected topics from recent fixed-order and resummed
calculations.  It focuses on how resummed perturbation theory has been used to
deduce the structure of nonperturbative corrections,
and to provide a framework with which to address the transition
from short- to long-distance dynamics in QCD.  
\end{abstract}

\section{Introduction}

Quantum chromodynamics (QCD) serves both as tool and background for new 
physics searches  \cite{snow}.
Control over QCD corrections is crucial for the interpretation of
any experiment in which hadronic matter
appears in the initial or final states, or even (as in the measurement of
the muon's magnetic moment) only in virtual states.  In addition,
QCD  is fascinating in its own right as a quantum field 
theory.
It combines strong and weak coupling dynamics in any high energy experiment
that is sensitive to momentum transfers much beyond 
the scale of the strong coupling, $\Lambda_{\rm QCD}$, and every experiment
that probes higher energies encodes the transition between
partonic degrees of freedom at short distances to hadronic degrees
of freedom at long.

This talk approaches these issues from the partonic,
short distance side, through the study of perturbation theory.  I will argue
that perturbative QCD has much to say about the structure of the larger
theory that contains it.  The discussion begins
with a sketch of basic concepts of perturbative QCD, 
followed by a review of some important recent developments at finite order.  
The extension of perturbative methods to
multiscale observables requires resummation, and I will describe how resummed
perturbation theory encodes nonperturbative information, with the operator
product expansion as the classic example.  Analogous reasoning is extended
to jet-related cross sections, and examples are given
of the interesting results that emerge from
such an analysis.  

\section{Some Concepts in Perturbative QCD}
\label{concepts}

\subsection{Asymptotic freedom and infrared safety}

The characteristic feature of QCD is its asymptotic freedom,
in which the coupling becomes weaker when it is measured
at shorter distances.   This remains a purely
abstract attribute, however, until we identify observable quantities
that can be expanded in the coupling with finite coefficients.
Such quantities are said to be infrared safe \cite{irs}.  A few cross 
sections,
such as the total and suitably-defined jet cross sections in
$\rm e^+e^-$ annihilation, are infrared safe, and
may be expanded as
\bea
{Q^2\;  \sigma_{\rm phys}(Q) }
&=&
{\sum_n c_n(Q^2/\mu^2)\; \alpha_s^n(\mu) + {\cal O}\left({1\over Q^p}\right)}
\nonumber\\
&=& {\sum_n c_n(1)\; \alpha_s^n(Q) +   {\cal O}\left({1\over Q^p}\right)}\, ,
\label{irs}
\eea
where $p$ is some integer and $\mu$ is a renormalization scale.
In the second expression we have used the renormalization group invariance
of physical cross sections.

\subsection{Factorization}

Cross sections for processes with hadrons in the initial state are not by
themselves infrared safe.  Nevertheless, if they involve
  heavy particle production, or more generally  involve
a large momentum transfer, $Q$, then short- and long-distance
components of the cross section can be factorized \cite{fact},
usually into a convolution form,
\bea
Q^2\sigma_{\rm phys}(Q,m)
=
\omega_{\rm SD}(Q/\mu,\as(\mu))\, \otimes\, f_{\rm LD}(\mu,m) + {\cal 
O}\left({1\over Q^p}\right)\, .
\label{fact}
\eea
In this expression, $\otimes$ represents a convolution, most
familiarly in terms of partonic momentum fractions, and $\mu$
is a factorization scale, separating the 
short-distance
dynamics of the hard subprocess, in $\omega_{\rm SD}$, from
the long-distance dynamics in $f_{\rm LD}$.  $f_{\rm LD}$ depends on
various infrared, often nonperturbative scales, denoted $m$.
For inclusive lepton-hadron deep-inelastic scattering (DIS),
in which $Q$ is the momentum transfer, and 
for various other inclusive and semi-inclusive hadronic hard-scattering cross 
sections, $f_{\rm LD}$
represents products of parton distribution functions.  For single-particle
inclusive cross sections, it includes fragmentation functions.
Generally speaking, `new physics' is part of the
short distance functions $\omega_{\rm SD}$, while the long-distance $f_{\rm 
LD}$'s
are universal among processes, making predictions possible 
on the basis of hypothetical couplings to new particles or rare processes at 
short distances.
Factorizations also apply to many scattering amplitudes and forms factors,
of which those involving the decay of B mesons have attracted
interest recently \cite{bfact}.

Eq.\ (\ref{fact}) applies not only to factorized cross sections with
hadrons in the initial state, but also to many
infrared-safe jet cross sections, for which the scale
$m$ is large enough to be perturbative, but is
still much smaller than the highest scale $Q$: $Q\gg m\gg\Lambda_{\rm QCD}$.
An example is the limit of `narrow jets' in $\rm e^+e^-$
annihilation, where $m$ may be taken as the mass of the jet \cite{muellerpr}.

\subsection{Factorization proofs and resummation}

The proofs  of relations like Eq.\ (\ref{fact})
are all based on the observation that short-distance
processes, organized within the function $\omega_{\rm SD}$, are 
quantum-mechanically
incoherent with  dynamics at much longer length scales.
This is the same observation that makes possible the operator
product expansion.  Extensions of Eq.\ (\ref{fact})
to jet cross sections and various decay and elastic
amplitudes depend on the further observation
that there is a mutual incoherence in the dynamics of particles
receding from each other at the speed of light.
Often, these factorization relations hold to
all orders in perturbation theory and to all
powers of $\ln (\mu/Q)$ \cite{factproof}.  Interest in these 
issues has reemerged in analyses 
of high energy processes in the language of
effective theories \cite{scet}.

Whenever there is factorization, there is some form of evolution and
resummation.  To see why, we observe that a physical cross section
cannot depend on the factorization scale,
\bea
0=\mu{d\over d\mu} \ln \sigma_{\rm phys}(Q,m)\, .
\eea
 From this it follows that the variations of long-distance
and short-distance functions with respect to $\mu$ must
compensate each other, through a function of the variables
that they hold in common:
\bea
\mu{d \over d\mu}\, \ln f_{\rm LD}(\mu,m)= - P(\as(\mu)) 
= - \mu{d  \over d\mu}\, \ln \omega_{\rm 
SD}\left(Q/\mu,\alpha_s(\mu)\right)\, .
\label{pcalc}
\eea
In deep-inelastic scattering, and other inclusive hard-scattering cross
sections, where $\omega_{\rm SD}$ and $f_{\rm LD}$ are connected by a 
convolution
in momentum fractions, the functions $P(z,\as)$ are 
the DGLAP splitting functions \cite{dglap}.

\begin{figure}[h]
%\vbox{\vskip 1 true in}
\centerline{\epsfxsize=7cm \epsfysize=9cm \epsffile{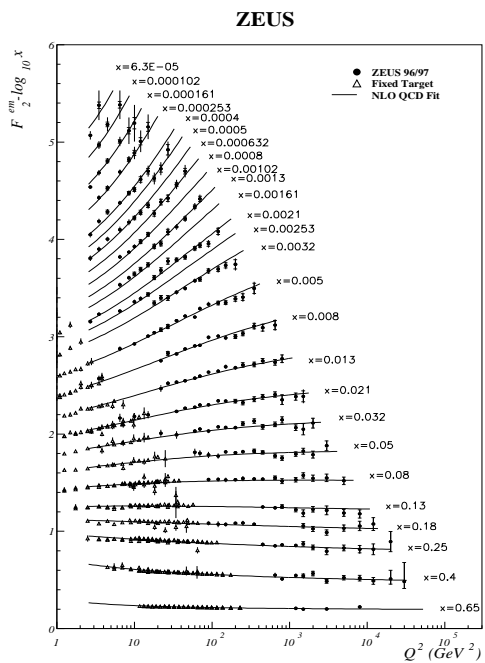}}
%\centerline{\epsfxsize=10cm \epsffile{scaling.eps}}
\caption{Data for $F_2(x,Q)$ with QCD fit based on evolution, from 
Ref.\ {\protect \cite{zeusf2}}.}
\label{f2fig}
\end{figure}

By the same token, wherever there is evolution, there is a resummation
of perturbation theory,
generated by solving the evolution equation.  Again thinking of
DIS, we may fix $\mu = Q$ in Eq.\ (\ref{fact}), compute $\omega_{\rm SD}$ 
as a power
series in $\as(Q)$, and derive the
$Q$-dependence of the cross section by solving (\ref{pcalc}) for
the functions $f_{\rm LD}(Q,m)$:
\bea
\ln \sigma_{\rm phys}(Q,m) &\sim& 
\omega_{\rm  SD}(1,\alpha_s(Q))\, 
\nonumber\\
&\ & \hspace{5mm} \otimes\, \exp\left\{  \int_q^Q {d\mu'\over \mu'}
P\left( \alpha_s(\mu')\right) \right\}\,  f_{\rm LD}(q,m)\, .
\eea
The most impressive successes of this approach are in observables
with a single hard scale.  Inclusive DIS is the exemplary case,
as illustrated in Fig.\ \ref{f2fig} for the structure function $F_2(x,Q)$,
which is proportional to the cross section 
to produce an hadronic final state of total invariant
mass squared equal to $Q^2(1-x)/x$, at momentum transfer $Q$.

\begin{figure}[h]
%\vbox{\vskip 1 true in}
\centerline{\epsfxsize=9cm \epsffile{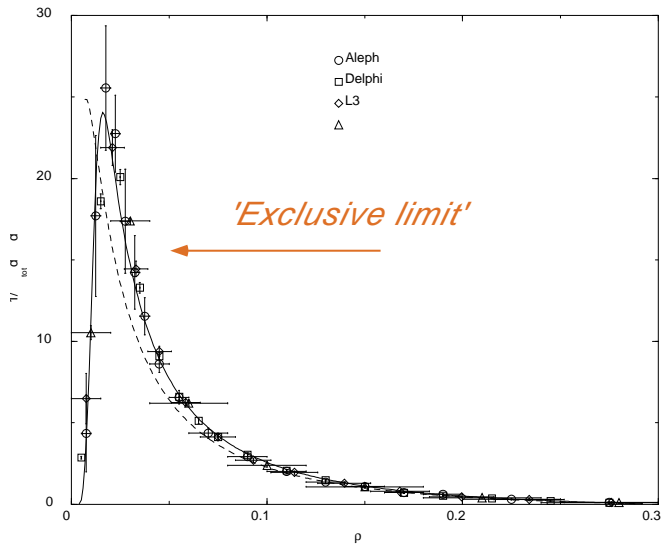}}
\caption{Heavy jet distribution at the Z pole, from Ref.\ {\protect 
\cite{KorTaf}}. The dashed line is resummed perturbation
theory, and the solid line a fit with a shape function.}
\label{hjm}
\end{figure}

As mentioned above, however, only a limited number of observables
depend on a single hard scale.  A more typical example is
the distribution of $\rho=(m_J/Q)^2$, from $\rm e^+e^-$
annihilation dijet events, with $m_J$ the mass
of the heavier jet in the 
center-of-mass (c.m.).
The experimental distribution in $\rho$ at the Z pole  
is illustrated in Fig.\ \ref{hjm}, taken from
\cite{KorTaf}.   In this case, the two scales are the
c.m.\ energy, $Q=\sqrt{s}$ and the mass of the `heavy' jet.
The data is shown as a function of $\rho=m_J^2/Q^2$, and
even resummed perturbation theory
(next-to-leading logarithm in $m_J/Q$) is not adequate to follow it.
It is necessary to include power  corrections in the
lighter scale: $1/m_J^p$ (as organized in a
shape function, see below), which become more and
more  important
in the `exclusive  limit' of light jet masses.  Notice that most 
events are in this range; going one  step beyond the fully inclusive cross
section, we are required to find a description with nonperturbative
as well as perturbative input, even though the observable in
question is infrared safe.   As we decrease $m_J$, nonperturbative
effects vary continuously from a few to tens of percent.  
Thus, within this single set of data, at a single energy, we
can select events whose formation is sensitive to 
an adjustable mix of perturbative
and nonperturbative dynamics.  This is an
ideal testing-ground for quantum field theory.

I shall discuss below recent attempts to organize 
nonperturbative corrections to infrared safe cross
sections starting with 
perturbation theory, in a manner similar to factorization
theorems like Eq.\ (\ref{fact}).
To do so with confidence,
we should attempt to organize perturbation theory to all orders
in the coupling, the level at which we understand Eq.\ (\ref{fact}).
First, however, let us touch on some of the very substantial progress
of the past few years that is based on finite-order calculations.

\section{Toward a Two-Loop Phenomenology}

Fig.\ 3 is a cartoon of the progress
of fixed-order calculations in perturbative QCD over the past 
three decades, with tree amplitudes under control 
around 1980, one-loop amplitudes around 1990, and
two-amplitudes since 2000 (early two-loop advances were reported
\cite{early2loop}, with recent reviews in \cite{2looprevs}).  
In the years following their calculation, 
amplitudes at the tree and one-loop level were
implemented into factorized cross sections like
(\ref{fact}).  With the hard-scattering computed
at tree level, the cross section is termed `leading order' (LO).
At one loop, it is `next-to-leading order' (NLO), and so on.  Now
we are at the dawn of an age of NNLO.  It is still
the dawn, because loop diagrams suffer from various infrared divergences,
which must be controlled by adding gluon emission diagrams
and constructing useable, factorized hard-scattering functions.
It is a big step from a set of (dimensionally) regulated amplitudes
to a factorized cross section (\ref{fact}) that can
be compared to experiment.  Still, with the two-loop
amplitudes in hand, the goal of NNLO 
phenomenology is certainly attainable.

\bigskip

\begin{figure}[h]
%\vbox{\vskip 1 true in}
\centerline{\epsfxsize=9cm \epsffile{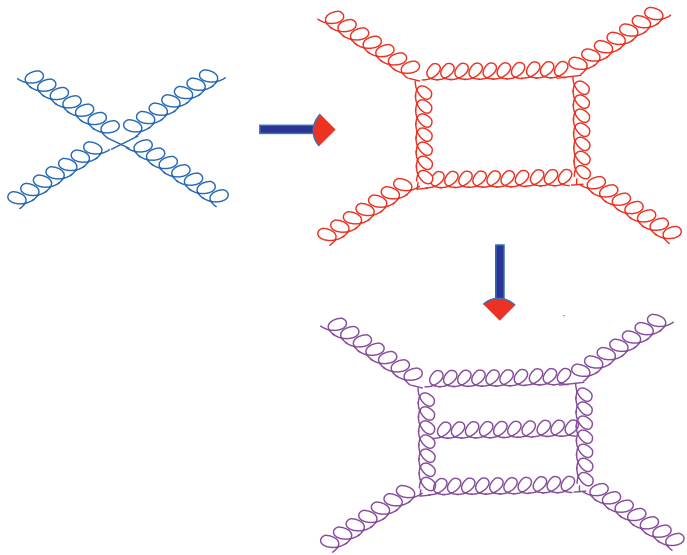}}
\label{cartoon}
\caption{Progress in low-order calculation.}
\end{figure}

Why go to the trouble of these very challenging calculations?
With $\as(Q) \sim 0.1$ at $Q\sim 100$ GeV,
computations to order $\as^2$
can reach the accuracy of one percent, an admirable goal
in its own right for the strong interactions.  At the
same time, accuracies of a few percent will enable
a new class of investigations of QCD,
by covering kinematic regions where nonperturbative
power corrections vary from a few percent to tens of percent,
as we have seen above.  Beyond this, one percent  accuracy will 
make possible a new generation of determinations of the strong coupling
itself, the uncertainty in whose precise value is a major 
limitation on our ability to extrapolate the standard
model into the realm of new physics \cite{burrows}.
Recent mileposts on this road include the calculation
of ${\cal O}(\as^3)$ 3-jet amplitudes
in electron-positron annihilation \cite{Garland3jet}.

For factorizable jet and other cross sections at
$n$ loops in the hard scattering, it is desirable
to have $n+1$ loops in the splitting function $P(z,\as)$,
in Eq.\ (\ref{pcalc}).  Progress toward   $P(z,\as)$ to $\alpha_s^3$
has included numerical estimates \cite{moments},
leading to studies of deep-inelastic scattering at this
level \cite{nnlo},
as well as a growing development of the methods necessary for full 
analytic expressions \cite{mvvcfb} at three loops.
The coming decade should see the 
full realization of the second step of Fig.\ 3.

The long-distance components of factorized cross sections are
parton distributions (PDFs) and fragmentation functions.  
Unpolarized parton distributions implemented 
with NLO calculations of hard scattering have
reached a stage of refinement over large
ranges of parton fraction $x$.  These refinements
are of great significance, because they are a limiting
factor in our ability to extrapolate known cross
sections to the scales at which new physics signals
may appear.  This said, there have still been `surprises',
in just the past few years, such as the unexpected
behavior of sea antiquarks as measured by 
experiment E866 at Fermilab \cite{e866}, and promptly
incorporated into the following round of parton  
distribution functions \cite{cteq6,mrstrecent}.  

Perhaps the most important recent development in the technology
of PDFs is the effort to quantify their uncertainties
\cite{uncert}.  This is a daunting task, because
any set of PDFs for all of the quarks, antiquarks and
the gluon requires the specification of a substantial
set of parameters.  Generally, in a `global' fit, these
parameters are fixed by comparison with a set of
existing data.  Any estimate of the uncertainty in a calculated
process depends in a unique way on all of
these parameters, each of which is sensitive to both
statistical and systematic errors in the original 
data set.   What we are interested in is not so much
the uncertainties in the PDFs themselves, but what these
uncertainties mean for cross sections that we
would like to calculate.   To determine this, it
is necessary to specify, not just a `best fit' for the
PDFs, but a set of degrees of freedom accompanied by 
correlated error estimates.  With this information,
it is possible to create an ensemble of PDFs, selected
according to their statistical likelihood.  
Each element of such an ensemble provides an
independent calculation of a factorized cross section
(\ref{fact}), and the distribution of these results
determines the desired uncertainty.  Needless to say,
this procedure is computationally-intensive in all
of its variants, depending on the choices of parameterization,
the analysis of parameter uncertainty, and the implementation
of the ensemble-creation process.
The rewards, however, are substantial, and the
effort is mandatory for extrapolation to
the range of the highest-energy hadron accelerators.

\section{Resummation}

Fixed-order calculations work best for single-scale 
observables, with the jet cross sections in hadronic
collisions offering a fine example, both of the
range of applicability, and the uncertainties involved
\cite{tevajet}.  The cross section for the production of bottom quarks
is an example involving more controversy, and 
possible recent progress \cite{bcross}.   Any finite-energy calculation,
however, leaves the nagging question of what happens
at yet higher order.
For single-scale inclusive cross sections, the
comparison of LO with  NLO and NNLO, when available,
can give confidence or shake it.  
(See, for example, recent NNLO calculations of inclusive
Higgs production at hadron colliders \cite{nnlohiggs}.)
Nevertheless, there
are certain general features of perturbative corrections
that demand study at all orders, even in one-scale
cross sections, and in most cases two-scale cross sections
cannot be controlled at all without a classification and
summation of such corrections at all orders.  Examples of
these two possibilities are threshold and
$Q_T$ resummation.

\subsection{Threshold resummation}

Threshold resummation addresses a specific set of corrections
at all orders in perturbation theory for factorized cross sections.
Consider a generic factorized cross section for the inclusive  production
of an observable final-state object $F$ of invariant mass $Q$, written in
terms of partonic fractions as
\bea
Q^2\sigma_{AB\rightarrow F}(Q)
=
  {f_{a/A}(x_a)} \otimes  f_{b/B}(x_b)
\otimes  \omega_{ab\to F+X}\left(z={Q^2\over x_ax_b S}\right)\, ,
\label{hadfact}
\eea
with the $f_{c/H}$ PDFs for parton $c$ in hadron $H$.
The hard-scattering cross section depends on variable $z$,
which is the ratio of $Q^2$ to the squared invariant mass 
of the pair of partons that initiate the process $ab\to F+X$,
where $X$ denotes additional radiation in
the inclusive final state.
Of special interest is the limit $z\to 1$, which
is called partonic threshold.  At partonic threshold,
the partons $a$ and $b$ have only enough energy to
produce $F(Q)$, with nothing left over for radiation.
This is a special limit in field theory, because a hard
scattering favors processes with copious radiation.  The
consequence of this mismatch is that $z=1$ is associated
with singular but integrable distributions, such as 
\bea
\omega_{ab\to F}^{(r)}(z) &\sim& 
\left({\alpha_s\over\pi}\right)^r {1\over r!}
\; \left[{\ln^{2r-1}(1-z)\over 1-z}\right]_+\, ,
\label{2kminus1}
\eea
where a plus distribution $[g(x)]_+$ with $g(x)$ 
singular at $x=1$ is defined
(like a delta function) by its integrals with smooth
functions:
\bea
\int_y^1 dx\, [g(x)]_+\, f(x) = \int_y^1\, dx\, g(x)\, [f(x)-f(1)] - 
f(1)\, \int_0^y dx\, g(x)\,  .
\eea 
In Eq.\ (\ref{hadfact}), the singular distributions are in
$\omega_{ab \to F+X}(z)$, which has the information
on perturbative radiation, and the smooth functions are
the PDFs.  While all of the plus distributions are
integrable, they become more and more singular at
higher orders, and it is natural to try to estimate their
overall effect beyond the lowest orders.

We are able to resum singular plus distributions because
at partonic threshold the hard-scattering process
returns to the elastic limit that we have referred to
above, in which final state radiation is grouped
into well-collimated jets, associated either with
the fragments of the incoming hadrons, or with
a set of nearly light-like outgoing jets.  Such
a momentum configuration is illustrated by Fig.\ 4
for the production of two jets.  

\begin{figure}[h]
%\vbox{\vskip 1 true in}
\centerline{\epsfxsize=8cm \epsffile{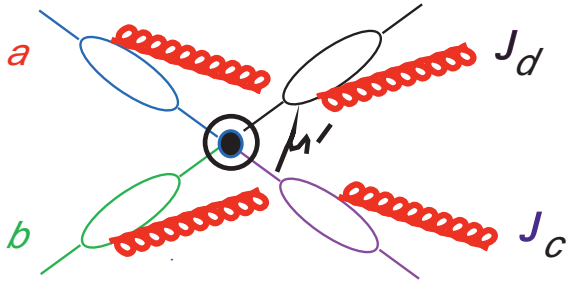}}
\label{refact}
\caption{Elastic kinematics and refactorization scale
at partonic threshold.}
\end{figure}

At partonic threshold, we may implement the sequence of reasoning
that we described in Sec.\ \ref{concepts}.  
Quanta are part of well-collimated jets, moving relatively at
the speed of light, or are very soft.  We may (re)factorize
the partonic cross section in this limit, 
as indicated in Fig.\ 4 by the circle and `refactorization scale',
$\mu'$, which leads to new
evolution equations, and hence to resummation.
For the hard scattering of Eq.\ (\ref{hadfact}), we find 
expressions whose expansion generates the terms identified
in Eq.\ (\ref{2kminus1}) to all orders in perturbation theory 
\cite{threresum}.
The resummation is most conveniently exhibited in
the space of Mellin moments of $\omega_{ab\to F+X}$,
where they exponentiate as
\bea
\int_0^1 dz z^{N-1} \omega_{ab\to F+X}(z)
&\ & \nonumber\\
&\ & \hspace{-30mm} \sim
\exp\left[\,  -\sum_{i=a,b}\int_0^1 dz{z^{N-1}-1\over 1-z}
\int_{(1-z)^2Q^2}^{Q^2}{d m^2\over m^2} A_i(\alpha_s(m))\;  \right]
\nonumber\\
&\ & \hspace{-30mm} \sim \exp\left[\; \sum_{i=a,b}\int_{Q^2/N^2}^{Q^2}{d 
m^2\over m^2} A_i(\alpha_s(m))
\, \ln\left( {Nm\over Q}\right) \; \right]\, ,
\label{llth}
\eea
in terms of anomalous dimensions
$A_i(\alpha_s)=C_i(\as/\pi)+\dots$, 
with $C_i=C_F$ for $i=q,\bar q$ and $C_A$ for gluons.
Eq.\ (\ref{llth}) generates the leading logarithms at each order; the complete 
expression involves process-dependent nonleading logs.  
At each order in the exponent, one of the logarithms can
be evaluated explicitly, as shown, but in the remaining integral
the strong coupling runs down to arbitrarily low scales.
We will return to this intriguing feature of the resummed cross
section below.  There are various ways of dealing with this
low-scale limit \cite{pvres,minres,beco,kid}, and although the
issue is far from settled, the bottom line seems to be
that the effect of singular distributions at partonic
threshold is a modest enhancement, along with a reduction
of dependence on the factorization scale.

\subsection{Color flow in resummation and at fixed order}

The expression (\ref{llth}) involves
radiation that is nearly collinear to the incoming and
outgoing partons at the hard scattering, and is completely
incoherent among these sets of lines.  
It organizes the leading logarithms  in $N$, and a 
portion of subleading corrections.
Beyond leading logarithm, however,
the active partons also radiate soft gluons coherently,
in a manner that depends on the exchange of color
at short distances -- schematically, inside the dashed
circle labelled $\mu'$ in Fig.\ 4.
At this level, threshold resummation  has an 
interesting
interplay with the one- and two-loop calculations 
\cite{early2loop,catani,tejeda} 
represented
by Fig.\ 3.

In the resummed cross section, the effects associated with
coherent soft-gluon emission are single-logarithmic in the moment
space of (\ref{llth}),
\bea
\exp \left[\, \int_{Q/N}^Q {d m \over m}\; \lambda^{\rm (f)}(\alpha_s(m))\, 
\right]\, ,
\eea
where the $\lambda$'s are eigenvalues of a set of anomalous dimension
matrices, which can be determined by separation of variables as in
Eq.\ (\ref{pcalc}).  They are matrices because the short-distance
function of Fig.\ 4 (inside the circle)
can be thought of as a vector in the space of color exchange.  For example,
when the active partons are all gluons, there are as many as eight 
independent color exchanges at short distances.  As $\mu'$ changes, gluons
that are exchanged near scale $\mu'$ enter or leave the hard scattering.
These effects are reflected in the NLO amplitudes of Fig.\ 3,
and influence the infrared structure of NNLO in an
essential way \cite{catani,tejeda}.

A striking example is the $gg\to gg$ anomalous dimension matrix, which
controls single logarithms in threshold resummation for gluonic
jet production \cite{kidonakiscolor}, with $N_c$ the number of colors:
\bea
{\small
\left(
                \begin{array}{ccccccccc}
                  T  &   0  & 0 & 0 & 0 & 0 & -\frac{U}{N_c} &\frac{T-U}{N_c} 
& 0  \vspace{2mm} \\ 
                  0  &  U & 0 & 0 & 0 & 0 & 0 & \frac{U-T}{N_c}& 
-\frac{T}{N_c}    \vspace{2mm} \\
                  0  &  0  &  T & 0 & 0 & 0 & -\frac{U}{N_c} &\frac{T-U}{N_c} 
& 0  \vspace{2mm} \\ 
                  0 & 0 & 0 &  \left( T+U \right) & 0 & 0 & \frac{U}{N_c} & 0 
&\frac{T}{N_c} \vspace{2mm} \\ 
                  0 & 0 & 0 & 0 &  U &0 &0 & \frac{U-T}{N_c} &-\frac{T}{N_c} 
\vspace{2mm} \\
                  0  & 0 & 0 & 0 & 0 &  \left( T+U \right) &\frac{U}{N_c} &0 
&\frac{T}{N_c} \vspace{2mm} \\ 
                 \frac{ T-U}{N_c}  &0 & \frac{T-U}{N_c} &\frac{T}{N_c} &0 
&\frac{T}{N_c} &2T &0 &0
\vspace{2mm} \\ 
                  -\frac{U}{N_c} &-\frac{T}{N_c} & -\frac{U}{N_c} & 0 
&-\frac{T}{N_c} &0 &0 &0 &0  \vspace{2mm} \\ 
                  0 &\frac{ U-T}{N_c} & 0 &\frac{U}{N_c} &\frac{U-T}{N_c} & 
\frac{U}{N_c} & 0 & 0 &2  U
                  \end{array} \right)\ ,
}
\nonumber\\
\eea
where $T=\ln\frac{1}{2}(1-cos\theta^*)$, $U=\ln\frac{1}{2}(1+\cos\theta^*)$ in
terms of the  c.m.\ scattering angle $\theta^*$.
The same matrix appears in explicit two-loop calculations of elastic 
gluon-gluon
amplitudes \cite{early2loop}, where it controls the single poles in 
dimensional regularization.
The all-orders nature of this relationship has been demonstrated in 
\cite{tejeda}.

\subsection{$Q_T$ and joint resummation}

\begin{figure}[t]
%\vbox{\vskip 1 true in}
\centerline{\epsfxsize=12cm \epsffile{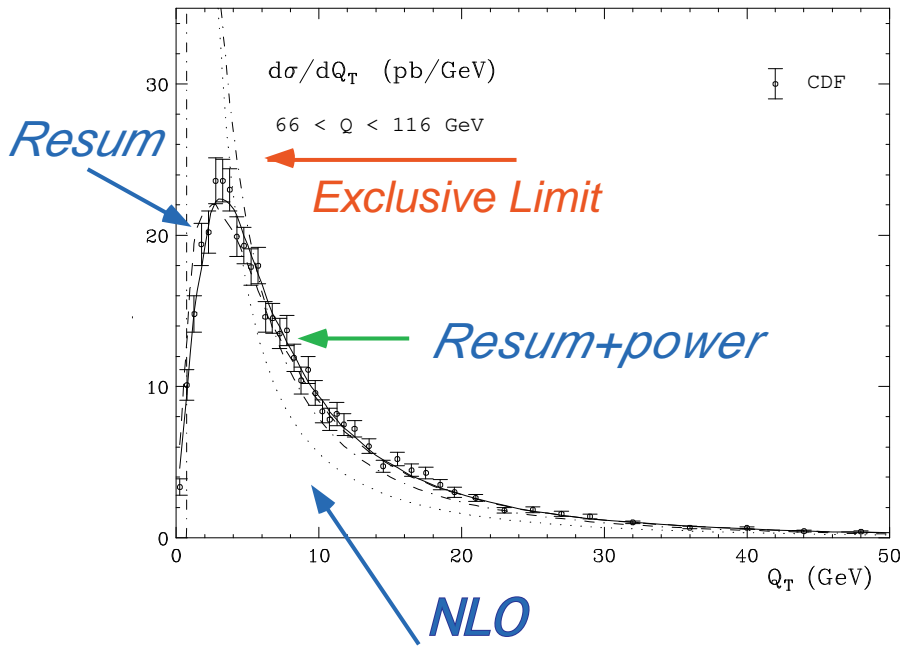}}
\caption{Jointly-resummed cross section 
$d\sigma/dQ_T$ for Z production at the Tevatron,
from Ref.\ {\protect \cite{kulesza}}.}
\label{Zqt}
\end{figure}

If threshold resummation strengthens our confidence in cross sections that
we can calculate to fixed order, there is a class of hadronic cross sections
for which resummation is absolutely necessary even to get started 
\cite{qtresum,qiuzhang}.  The most familiar of these are the hadronic 
electroweak
annihilation cross sections at measured transverse momentum ($Q_T$) for the
production of an  electroweak boson 
$B(Q)$ (virtual photon,
W, Z or Higgs, of mass $Q$).  In the underlying processes $q\bar{q}\to B(Q)$ 
or 
$gg\to B(Q)$ (the latter through a fermion loop),
the limit $Q_T\to 0$ is singular in much the same way as $1-z\to 0$ at
partonic threshold, Eq.\ (\ref{2kminus1}), with plus distributions in $Q_T^2$ 
up to $(1/Q_T^2)\, \ln^{2k-1}(Q_T^2/Q^2)$ at order $\as^k$.
Indeed, these two limits are closely connected and involve
the same underlying factorization, and the resummation necessary
for logarithms in transverse momentum is compatible 
 with  threshold resummation \cite{beco,liunif}.   
Again, the result for the factorized/resummed  cross
section exponentiates in transform space, in this case a combination of
Mellin ($N$) and Fourier ($b$, `impact parameter') transforms.  The
`jointly resummed' physical 
cross section is then approximated by a double
inverse transform,
\bea
{d\sigma^{\rm res}\over dQ^2dQ_T^2}
&\ & \sim\ \sum_a\ \sigma_0^{a\bar a \rightarrow B(Q)}\;
\int_C {dN\over 2\pi i}\, \tau^{-N}\int_{C_b} {d^2b\over (2\pi)^2}\, {\rm 
e}^{i\vec
Q_T\cdot
\vec b}
\nonumber\\
&\ & \hspace{10mm} \times\ {\cal C}_a(Q,b,N)\, 
{\rm e}^{E_{a\bar a}(N,b,Q)}\, 
{\cal C}_{\bar a}(Q,b,N)\, ,
\label{jores}
\eea
where the PDFs are included in the factors
\bea
C_a(Q,b,N,\mu)&=& \sum_j C_{a/j}\left(N,\alpha_s(\mu)\right)\, 
f_j\left(N,Q/(N+bQ)\right)\, .
\eea
The leading logarithmic corrections are of much the same form as
in threshold resummation, Eq.\ (\ref{llth}),
\bea
E_{a\bar a}(N,b,Q) \sim 2\; \int_{[Q/(N+bQ)]^2}^{Q^2}{d m^2\over m^2} 
A_a(\alpha_s(m))\,
  \ln\left({m\over Q}  \right)\, ,
\label{jointexp}
\eea
in terms of the same anomalous dimensions $A_a(\as)$.  A resummed cross 
section
calculated from (\ref{jores}) is compared to Tevatron data 
for Z production in Fig.\ \ref{Zqt}.
Similar good fits are found on the basis of $Q_T$-resummation alone
\cite{qtresum,qiuzhang}.  We note in the figure a general resemblance to the
heavy jet mass distribution of Fig.\ \ref{hjm}, where now the exclusive
limit is at $Q_T=0$, which has the kinematics of quark-antiquark annihilation
without gluon radiation.  There is a need for modest
\cite{qiuzhang,kulesza} but nonzero  `power' 
corrections, as shown in the figure.

\section{Power Corrections and Event Shape Functions}

Examining Figs.\ \ref{hjm} and \ref{Zqt} we see that for differential 
distributions,
power corrections are necessary to bring even resummed cross sections
into agreement with data in the
exclusive limit.  In the
range where the kinematics are far from elastic, fixed-order perturbation
theory is enough, but as we climb the curves of these figures toward
the elastic limit, the cross section rises and we move into a region
where first resummation, and then nonperturbative corrections dominate.
The theory itself knows no strict boundary between perturbative
and nonperturbative regimes.  It is therefore natural 
to analyze perturbation theory to see if it gives signals of its
own incompleteness, and to seek hints on how to supplement
perturbative predictions with nonperturbative information.
The ultimate goal is to construct
a single theory that bridges the gap between short- and long-distance
dynamics.

\subsection{The operator product expansion}

The approach that I will describe below begins with 
the simplest one-scale problem, the total
cross section in $\rm e^+e^-$ annihilation \cite{Mueller8592}.
The total cross section
is related by the optical theorem to the imaginary part of the
vacuum polarization for electroweak currents,
\bea
\pi(Q^2)=-{i\over 3Q^2}\; \int d^4x\; {\rm e}^{-iq\cdot x}
\langle 0|T j^\mu(0)j_\mu(x)|0\rangle\, ,
\eea
whose behavior at large $Q^2\equiv q^2$ is determined by an operator product 
expansion: 
\bea
\langle 0|j^\mu(0)j_\mu(x)|0\rangle 
&=& {1\over x^6}\; {\tilde C}_0\left(x^2\mu^2,\as(\mu)\right) \nonumber\\
&\ & \ 
\hspace{-20mm}+{1\over x^2}\; 
{\tilde C}_{F^2}\left(x^2\mu^2,\alpha_s(\mu)\right)
\langle 0|F_{\mu\nu}F^{\mu\nu}(0)|0\rangle +\dots\, .
\eea
Here, I have shown only the contributions of the leading (identity) 
operator and of the gluon condensate $F^2$.  Neglecting masses,
perturbation theory alone would give only the leading term,
even though the full theory demands both (and more).  The question
we ask is whether we could have discovered the need for
the nonleading operators from perturbative QCD alone.  

\begin{figure}[h]
%\vbox{\vskip 1 true in}
\centerline{\epsfxsize=8cm \epsffile{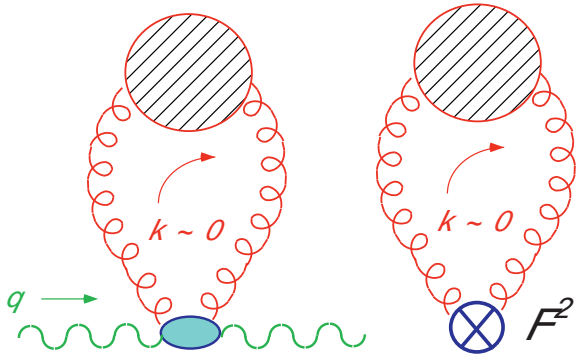}}
\caption{Sources of infrared sensitivity in 
perturbation theory for $C_0$ (left) and $\langle F^2\rangle$ (right).}
\label{pifsquare}
\end{figure}

The answer is yes, and the reasoning is illustrated in Fig.\ \ref{pifsquare},
which shows graphical configurations where loop momenta $k$
are vanishingly small for the vacuum polarization and for
the gluon condensate treated as a local operator.  Diagram-by-diagram,
each such configuration is a tiny `corner' of loop momentum space,
where one or more loops has low momentum,
giving a finite contribution to the infrared safe integrals.
For example, if the loop momentum is required to
have $k^2<\kappa^2$, for some cutoff $\kappa$, this
contribution is suppressed by $(\kappa^2/Q^2)\, \alpha_s(\mu)$ compared
to the full diagram.  It is thus indeed power suppressed.
Using gauge invariance and the renormalization group, however,
we can show that summing over the self-energies indicated in
the figure has the effect of making the coupling run and 
of giving an even greater suppression.  For the
vacuum polarization, we find 
momentum-space integrals of the general form
\bea
C_0^{\rm IR}(Q,\kappa)
 &=&  \frac{1}{Q^4}\, H(Q)\, \int_0^{\kappa^2} dk^2 k^2\;  \alpha_s(k^2) 
\nonumber \\
\nonumber \\
 &\ & \hspace{-20mm} = \frac{1}{Q^4}\, H(Q)\, \int_0^{\kappa^2} dk^2 k^2
\ {\alpha_s(Q^2) \over 1 + \left ({\alpha_s(Q^2)\over 4\pi}\right ) 
\beta_0\ln(k^2/Q^2)}\, ,
\label{irint}
\eea
with $H(Q)$ an infrared safe function free of all contributions
where a single loop momentum vanishes.  $H(Q)$ represents the lower,
contracted part of the diagram on the left of Fig.\ \ref{pifsquare}.
Up to this factor, the expression for the right-hand
figure, corrections to the condensate, is identical.
A generalization of
Eq.\ (\ref{irint}) is valid to all orders in perturbation theory,
and controls all logarithms in the loop momentum $k$, up to
the next power, $Q^{-6}$ \cite{Mueller8592}.  We will
call an expression at this level of accuracy, all logs
at a given nonleading power, an `internal resummation' at that power.  

From (\ref{irint}) we
learn first, that the internally-resummed contributions 
are singular for very small $k^2$, from high order diagrams.
Correspondingly, it is easy to verify that the $n$th order of the reexpansion
of Eq.\ (\ref{irint}) in $\alpha_s(Q)$ is proportional to $n!$.  We also learn
that these infrared-sensitive contributions in the vacuum
polarization generate exactly the perturbative expansion of
the gluon condensate.  We conclude that the perturbative vacuum
polarization requires the presence of a nonperturbative gluon condensate,
which should replace its misleading high-order infrared behavior.

The foregoing steps may be summarized as an `axiom of substitution',
in which the infrared-sensitive, nonconvergent portion of
an internally-resummed perturbative quantity is replaced by
a matrix element with the same perturbative infrared expansion:
\bea
C_{\rm PT} &\rightarrow& C_{\rm PT}^{\rm reg}\left 
(Q^2/\mu^2,\kappa/Q,\alpha_s(\mu)\right )
\nonumber\\
&\ & \hspace{10mm}
+\; {1 \over Q^4}\, C_{F^2}(Q,\kappa)\ \alpha_s 
\langle0|F^2(0)|0\rangle(\kappa) \, .
\eea
Here, $C_{\rm PT}^{\rm reg}$ is perturbation theory subtracted
in its infrared-sensitive limit.  
As anticipated, the nonconvergence of perturbation theory implies the need for
a new, infrared regularization.
For the vacuum polarization the cost is a new 
nonperturbative parameter ($\langle F^2 \rangle$), necessary to define the
theory at the level of $1/Q^4$ relative to the leading behavior.
But, because it is
implicit in perturbation theory, this new parameter is at the same time a 
reward of
the  analysis.

\subsection{Power corrections in semi-inclusive cross sections}

The viewpoint described above for the operator product expansion
has been implemented, in various ways, to a variety of infrared safe
cross sections and hard-scattering functions  
\cite{irr}.  
Of particular interest are resummed hard scattering functions like 
Eqs.\ (\ref{llth}) and (\ref{jointexp}),
which exhibit the same integral over the scale of the coupling as for
the vacuum polarization (\ref{irint}).  
The application of these methods to event shape
cross sections \cite{irrdiff}
sheds light on the transition from short- to long-distance
dynamics in the formation of final states in QCD.  
This section will sketch the ideas behind these applications, 
and also touch on internal resummation in this context.  
Such questions are most easily addressed for
event shapes of two-jet cross sections in $\rm e^+e^-$ annihilation,
although their application is much wider \cite{irr3j}.

We consider inclusive cross sections defined by
event shapes,   $S(\{p_i\})$, 
\bea
{d \sigma \over d \cs}
=
{1\over 2Q^2}\ \sum_n \int_{PS(n)} |M_n(\{p_i\})|^2 \delta \left ( 
S(\{p_i\})-\cs\right) \, ,
\eea
at c.m.\ energy $Q$, computed from QCD matrix elements $M_n(\{p_i\})$
integrated over $n$-particle phase space, $PS(n)$.
This event shape cross section is infrared safe 
if $S(\{p_i\})$ is a smooth function of final state momenta, $p_i$,
that is unchanged by soft parton emission
and collinear rearrangements, as illustrated in Fig.\ 7:
\bea
S(\dots p_i \dots p_{j-1},\alpha p_i,p_{j+1}\dots ) =
S(\dots (1+\alpha) p_i \dots p_{j-1},p_{j+1}\dots ) \, .
\eea 
This condition makes it possible to compute the cross
section perturbatively,
because it deemphasizes the long-time evolution of the system,
over which the nearly degenerate states of Fig.\ 7 mix.

The study of event shapes in QCD, and more generally energy flow \cite{flow},
continues a long tradition in field theory.
We may draw a not too far-fetched analogy
between how, late in the nineteenth century Poynting discovered the rules
for energy and momentum flow in classical electrodynamics
by considering the slow discharge of a condenser
 \cite{Poynting}, and how, late in the twentieth century,
the short-distance structure of QCD was revealed in
the rapid neutralization
of a quark-antiquark color dipole produced in
$\rm e^+e^-$ annihilation \cite{hansonetal}. 

\begin{figure}[h]
%\vbox{\vskip 1 true in}
\centerline{\epsfxsize=7cm \epsffile{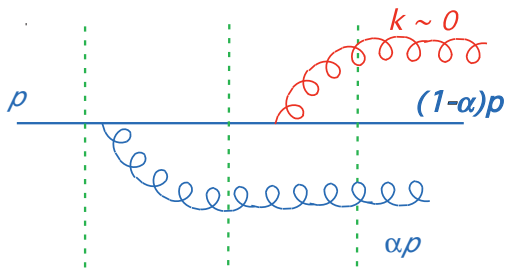}}
\label{rearr}
\caption{Soft emission and collinear rearrangements.}
\end{figure}

We restrict ourselves here to a specific set of weights, appropriate to dijet 
events in annihilation cross sections \cite{berkucste2}, 
\bea
{\cal S}_a(n) 
= {1\over Q}\sum_{{\rm all}\ i\in n}\ k_{iT}\; \e^{-|\eta_i|(1-a)}\, ,
\eea
where the transverse momentum $k_{iT}$ and pseudorapidity
$\eta_i$ of each particle momentum
$k_i$ is defined relative to the axis that minimizes 
the specific function ${\cal S}_0(n)$
for state $n$.
The variable $a$ takes on any value less than two.
The particular choice $a=0$ is ${\cal S}_0=1-T$ with $T$ the thrust,
while $a=1$ is known as the jet broadening \cite{cttw}.

For these dijet event shapes, sensitivity to long times can be adjusted by
studying the Laplace transform \cite{irrdiff,cttw},
\bea
\tilde{\sigma}_a (N) = {1\over \sigma_{\rm tot}} \int^1_0 d \cs_a\, \e^{\; 
-N\, 
\cs_a}\ {d 
\sigma \over d \cs_a}\, .
\label{Laplace}
\eea
Choosing $N$ large, we select states
with small $\cs_a$, which are  
sensitive to long-time behavior (roughly, times of order
$1/S_aQ$), corresponding to low jet masses in
dijet events.  
As illustrated by Fig.\ \ref{hjm}, however,
power corrections are important at low jet masses.
We thus ask what the reasoning described above for the operator
product expansion can tell us about the
transformed cross sections, especially at large $N$ in Eq.\ 
(\ref{Laplace}).

At low values of ${\cal S}_a$, these differential event shape cross sections 
show
the same double logarithmic behavior discussed above
for threshold resummation, and 
obey a similar resummation in Laplace transform space \cite{berkucste2,cttw}, 
which  is given to leading logarithm, accompanied by effects
of the running coupling, by
\bea
\tilde{\sigma}_a (N,Q)
&\sim&
\exp\; \Bigg\{\, 2\,   
\int_{0}^{Q^2} {dp_T^2\over p_T^2}\; A_q\left(\alpha_s(p_T)\right)
\,
\int_{p_T^2/Q^2}^{p_T/Q} {dy \over y} \nonumber \\
&\ & \hspace{30mm} \times
\left[ \e^{ -N\, (p_T/Q)^a\, y^{1-a} } -1\right] \Bigg\}\, ,
\label{resumcs}
\eea
in terms of the same Sudakov anomalous dimension as above, $A_q(\as)=$ 
$C_F(\as/\pi)+\dots$.   

\begin{figure}[h]
\centerline{\epsfxsize=9cm \epsffile{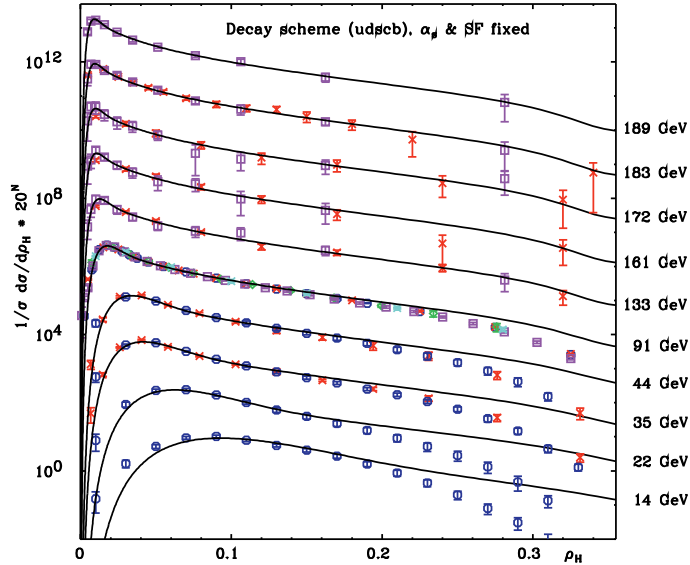}}
\caption{Event shape-based fit to the heavy jet mass at various energies, from Ref.\ 
{\protect \cite{GardiR}}.} 
\label{thrust}
\end{figure}

We can use Eq.\ (\ref{resumcs}) to 
suggest the structure of power corrections in differential 
cross sections, as described above for the total annihilation cross section.
To do so, we split up the $p_T$ integral into `perturbative'
and `soft' ranges, separated at a factorization scale
$\kappa$ \cite{KSt99BKorS}.  If we expand the exponent of (\ref{resumcs}) in
the soft range, we find
\bea
\ln  \tilde\sigma_a(N,Q)
&=&    \ln  \tilde\sigma_{a,\pt}(N,Q,\kappa) 
\nonumber\\
 &\ & \hspace{-30mm}
+
\frac{2}{1-a}\
\sum_{n=1}^\infty\ \frac{1}{n\, n!}\, {\left(-{N\over Q}\right)}^{n}
\int_{0}^{\kappa^2} {dp_T^2\over p_T^2}\; p_T^n\; A\left(\alpha_s(p_T)\right)\
\left[ 1 - \left({p_T \over Q}\right)^{n(1-a)}\right]
\nonumber\\
&\ & \hspace{-15mm} \equiv
\ln \tilde\sigma_{a,\pt}(N,Q,\kappa) + \ln \tilde f_{\cs_a}(N/Q) + {\cal 
O}\left( 
{N\over Q^{2-a}}\right)\, .
\label{pcshape}
\eea
In the second line we organize the entire contribution
from the soft region into a single function $\tilde f_{\cs_a}(N/Q)$,
with corrections relatively suppressed by powers in $1/Q^{1-a}$
(assuming for now that $a<1$).  Taking the inverse transform
of (\ref{pcshape}) we arrive at 
an expression for the physical cross section as a convolution
of the perturbative cross section with what is known as
an event shape function \cite{KSt99BKorS},
\bea
{d\sigma \over d{\cs_a}} =
\int_0^{{\cs_a}\, Q} d\epsilon\; f_{\cs_a}(\epsilon)\; 
{d\sigma_{a,\pt}({\cs_a}-\epsilon/Q) \over
d{\cs_a}} +{\cal O}\left({1\over {\cs_a} Q^{2-a}}\right)\, .
\label{convol}
\eea
The event shape function $f_{\cs_a}$ is independent of $Q$,
so that a fit to data at $Q=m_Z$ is sufficient
to predict the differential cross section at any $Q$,
including all integer powers of $N/Q$ in moment
space \cite{KorTaf}, which translates to all integer powers of $1/\cs_a Q$ 
at fixed $Q$.  This is the kind of formalism necessary to 
confront data like those in Fig.\ \ref{hjm} above.
Analogous functions find important applications in the physics
of b quarks \cite{bshape}.  A recent
example appled to the heavy jet mass, closely
related to $a=0$, is shown in Fig.\ \ref{thrust}
from Ref.\ \cite{GardiR}. 
Beyond the $a=0$, the shape functions derived in this fashion for choices of 
$a,b<1$ possess a surprisingly
simple relation,
\bea
\ln\, \tilde f_{\cs_a}(N/Q) = {1-b\over 1-a}\; \ln\, \tilde f_{\cs_b}(N/Q)\, ,
\label{freln}
\eea
with corrections as in (\ref{convol}).  For 
$2>a>1$, on the other hand, we encounter the interesting situation
noted first in Ref.\ \cite{manwise}, where the leading
power correction becomes fractional, and larger than $1/Q$.

The next question we should ask, however, is whether
Eqs.\ (\ref{resumcs}) and (\ref{freln}) hold beyond
leading and (suitably generalized \cite{cttw}) next-to-leading logarithmic resummation,
and in particular whether these event shapes can be internally resummed
to all orders in $1/\cs_a Q$.
Rather than try to answer
this question for the full theory, we will follow Ref.\ \cite{KSt99BKorS}, and
consider the `eikonal' cross section, in which the
primary quark-antiquark pair and the partons
collinear to them are all replaced by nonabelian phase operators.
These are Wilson lines, represented by 
products of exponentials of the gauge
field ordered along lightlike paths,
\bea
W =  T\Bigg(  \; 
\Phi^{(q)}_{\beta_1}(\infty,0) \; \Phi^{(\bar q)}_{\beta_2}(\infty,0) 
\Bigg )\,  ,
\label{phiprod}
\eea
with
\bea
\Phi^{(i)}_{\beta _i}(\lambda,x)
=
{\cal P}\exp \left (-ig\int_0^\lambda d\lambda' \beta _i\cdot 
A^{(i)}(\lambda'\beta_i+x)
\right )\, ,
\eea
where $A^{(i)}$
is the gluon field in representation $i$.  In these terms,
the eikonal cross sections at fixed values of
event shapes $\cs_a$ may be written as a sum over states, $n$:
\bea
{1\over \sigma_{\rm tot}}\
{d \sigma^{\rm eikonal}\over d\cs_a}
=
\sum_n\  \delta\left(\cs_a-\cs_a(n)\right)\ 
 \langle 0  |W^\dagger \, 
  |n  \rangle\,  \langle n |
\, W\,  | 0  \rangle\, .
\eea
These cross sections have the advantage that many of their properties
may be analyzed to all orders in perturbation theory
using combinatoric methods \cite{webs}.
Here I will simply quote the result of such an analysis.

As in Eq.\ (\ref{resumcs}), the all-orders eikonal approximation
to the cross section
can be written in transform space in terms of a function in which 
$\as$ runs with an integration variable, which we may think of as 
a  transverse momentum,
\bea
\ln\; \tilde \sigma_a(N,Q)
=
\int_0^{Q^2} {d^2p_T\over p_T^2}\, 
{\cal A}_{{\cal S}_a}\left(p_T/\mu,Q/\mu,\as(\mu),N\right)\, ,
\eea
where the function ${\cal A}_{{\cal S}_a}$ is 
infrared safe and renormalization group invariant 
($d{\cal A}_{{\cal S}_a}/d\mu=0$).   As an example, consider
the case $a=1$, jet broadening, for which we can write
\bea
{\cal A}_{{\cal S}_1}\left(p_T/\mu,Q/\mu,\as(\mu),N\right) &\ &
\nonumber\\ &\ & \hspace{-45mm}
= \sum_n \int_{PS(n)}\
{\cal W}_n\left({k_i\cdot k_j\over \mu^2},{k_i\cdot \beta_1
k_j\cdot \beta_2\over\mu^2\beta_1\cdot\beta_2},\as(\mu)\right)
\nonumber\\
&\ & \hspace{-40mm}  \times
\delta^2\left(p_T-\sum_{i\in n} k_{i,T}\right)\,  \delta \left(\eta_{n}\right)
\ln(Q/p_T)\, \left[ \exp \left\{ -{N\over Q}\,  p_T\right\} -1\right]\, ,
\nonumber\\
\label{broadening}
\eea
where the $k_i$ denote particle momenta in final state $n$, and
where $\eta_n$ is the rapidity of the {\it total} momentum of
state $n$.  Boost invariance requires that the 
functions ${\cal W}_n$ \cite{webs} depend only on the arguments shown.
For other event shapes, such as the thrust, 
slightly more elaborate expressions are necessary.  In each case,
however, a meaningful factorization of the infrared, strong-coupling
dynamics from the perturbative region is possible.  Eq.\ (\ref{broadening}), 
although
valid for all orders and logarithms, applies only
to the eikonal cross section. 
Nevertheless, it appears that the way 
is open for the all-orders analysis of the 
relation between perturbation theory
and power corrections in event shape functions and
related parameterizations, beyond the  limitations of 
next-to-leading logarithm.

\section{Summary}

Short-distance, single-scale observables in QCD are reasonably well 
understood,
which is one reason why we have such confidence that QCD is a true theory
of the strong interactions.  The developing technology of
next-to-next-to-leading order, coupled with a sophisticated
approach to parton distribution uncertainties, should
make possible precision for selected observables at the level
of one percent.

At the same time, we are far from a full understanding
of multi-scale observables, which lead us
from perturbative to nonperturbative dynamics.
Power corrections and  event shape functions for 
infrared safe differential distributions
may be identified by an extension of the reasoning that leads to the operator
product expansion for the total
$\rm e^+e^-$ annihilation cross section.
By exploring concepts like internal resummation, 
 we may find
a perturbative window to the formation of final
states, and gain insight into the transition between
short- and long-distance degrees of freedom
in quantum chromodynamics.

\subsection*{Acknowledgements}

I thank the organizers of
TH2002 and of the 26{\it th} Johns Hopkins
Workshop for their invitations, and for
making possible my participation.  This work was supported in part by
the National
Science Foundation grant PHY0098527.

\end{document}